\documentclass[aps,showpacs]{revtex4}
\usepackage{amsmath}
\usepackage{amssymb}
\usepackage{graphicx}
\usepackage{epsfig}
\usepackage{color}
\usepackage{url}

\begin{document} 
\title{Numerical relativity with characteristic evolution, using six angular patches} 
\author{Christian Reisswig${}^{1}$, Nigel T. Bishop${}^{2}$, Chi Wai Lai${}^{2}$,
Jonathan Thornburg${}^{1}$, and Bela Szilagyi${}^{1}$}
\affiliation{${}^{1}$
  Max-Planck-Institut f\"ur Gravitationsphysik, Albert-Einstein-Institut,
  Am M\"uhlenberg 1, D-14476 Golm, Germany \\
${}^{2}$
  Department of Mathematical Sciences, University of South Africa,
  P.O. Box 392, Unisa 0003, South Africa}

\begin{abstract} 
The characteristic approach to numerical relativity is
a useful tool in evolving gravitational systems. In the past this has
been implemented using two patches of stereographic angular coordinates.
In other applications, a six-patch angular coordinate system has proved
effective.  Here we investigate the use of a six-patch system in
characteristic numerical relativity, by comparing an existing two-patch
implementation (using second-order finite differencing throughout)
with a new six-patch implementation (using either second- or fourth-order
finite differencing for the angular derivatives). We compare these
different codes by monitoring the Einstein constraint equations, numerically
evaluated independently from the evolution.  We find that,
compared to the (second-order) two-patch code at equivalent resolutions,
the errors of the second-order six-patch code are smaller by a factor of
about~2, and the errors of the fourth-order six-patch code are smaller
by a factor of nearly~50.
\end{abstract} 

\maketitle 


\section{Introduction}
\label{s-intro}
The characteristic, or null-cone, approach to numerical relativity 
is based on the Bondi-Sachs metric~\cite{bondi,sachs}, and has
been successfully implemented in the PITT code~\cite{hpn,roberto,cce,
mat,particle,fission} (and also in other
codes~\cite{ntb90,rdi1,rdi2,philiptoni,siebel1,siebel2,bartnik}). The PITT
code has the important property that all tests have shown it to be
long-term stable, for example in evolving a single black hole
spacetime~\cite{wobble}. However, there are problems with the computation
of gravitational radiation: it has been successfully computed in test
cases~\cite{hpn,cce} and also in scattering problems~\cite{mod}, but it has
not been possible to compute
gravitational radiation emitted in astrophysically interesting
scenarios, such as a star in close orbit around a black hole~\cite{particle,
mat2}.

This paper is a contribution towards the longer term goal of producing a
characteristic code that can reliably compute gravitational radiation in
situations of astrophysical interest, either within a stand-alone
code~\cite{particle,mat2}, or within the context of Cauchy-characteristic
extraction or Cauchy-characteristic matching~\cite{cce2}.
The first step in this process was to find a class of exact solutions to
the linearized Einstein equations~\cite{BS-lin}. These solutions are
written in terms of the Bondi-Sachs metric and continuously emit
gravitational radiation. Thus they provide an appropriate testbed for
validating further developments of the characteristic code.

In this paper, we investigate a strategy for improving the code. In
another context in numerical relativity~\cite{jthorn}, good results have
been obtained by coordinatizing the sphere by means of six angular patches,
rather than the two patch stereographic coordinates used in the PITT code.
One advantage of the six patch system is that inter-patch interpolation is
particularly simple because it is one-dimensional (since ghost zone points
lie on grid lines), leading to an expectation of reduced noise at the patch
interfaces. Further, this means that it is easy to go from second- to
fourth-order accurate finite differencing of angular derivatives, since the
interpolation order must also be increased and this is straightforward in
the one-dimensional case. Using the exact solutions of ref.~\cite{BS-lin}
as a testbed, we have computed and compared the errors obtained in (a) the
stereographic code, (b) the second-order six patch code, and (c) the
fourth-order six patch code. Note that, in the fourth-order case, the radial
and time derivatives are second-order accurate, so that overall the code is
second-order -- unfortunately, changing these derivatives to fourth-order
would be a major undertaking, because they are not confined to a single
subroutine. We found that, for equivalent resolutions, the errors of (b)
are a little smaller than those of (a), and that the errors of (c) are
much smaller than those of (a) by a factor of nearly 50.

As in the ADM formalism, four of the ten characteristic Einstein equations
are not used in the evolution but constitute constraints. We have
constructed and validated code that evaluates the constraints, and this
can be used as a tool to monitor the reliability of a computational
evolution.

The numerical computations presented in this paper were all performed
within the Cactus computational
toolkit~\cite{Goodale02a}~(\url{http://www.cactuscode.org}),
using the
Carpet driver~\cite{Schnetter-etal-03b}~(\url{http://www.carpetcode.org})
to support the multiple-patch computations.
The computer algebra results were obtained using Maple.

The plan of the paper is as follows. Sec.~\ref{s-back} summarizes background
material that will be used later. Sec.~\ref{s-6patch} describes our
implementation of the six patch angular coordinate system. Sec.~\ref{s-cons}
describes the constraint evaluation. Computational results are presented
in Sec.~\ref{s-res}, and are then discussed in the Conclusion,
Sec.~\ref{s-conc}.


\section{Background material}
\label{s-back}

\subsection{The Bondi-Sachs metric}
The formalism for the numerical evolution of Einstein's equations, in
null cone  coordinates, is well known~\cite{hpn,roberto,cce,mat,rai83,
bondi}. For the sake of completeness, 
we give a summary of those aspects of the formalism that will be used here.
We start with coordinates based upon a family of outgoing null
hypersurfaces.
We let $u$ label these hypersurfaces, $x^A$ $(A=2,3)$, label
the null rays and $r$ be a surface area coordinate. In the resulting
$x^\alpha=(u,r,x^A)$ coordinates, the metric takes the Bondi-Sachs
form~\cite{bondi,sachs}
\begin{eqnarray}
 ds^2  =  -\left(e^{2\beta}(1 + W_c r) -r^2h_{AB}U^AU^B\right)du^2
\nonumber \\
        - 2e^{2\beta}dudr -2r^2 h_{AB}U^Bdudx^A 
        +  r^2h_{AB}dx^Adx^B,
\label{eq:bmet}
\end{eqnarray}
where $h^{AB}h_{BC}=\delta^A_C$ and
$det(h_{AB})=det(q_{AB})$, with $q_{AB}$ a metric representing a unit
2-sphere embedded in flat Euclidean 3-space; $W_c$ is
a normalised variable used in the code, related to the usual Bondi-Sachs
variable $V$ by $V=r+W_c r^2$. As discussed in more detail below, we
represent $q_{AB}$ by means of a complex dyad $q_A$. Then,
for an arbitrary Bondi-Sachs metric,
$h_{AB}$ can then be represented by its dyad component
\begin{equation}
J=h_{AB}q^Aq^B/2,
\end{equation}
with the spherically symmetric case characterized by $J=0$.
We also introduce the spin-weighted field
\begin{equation}
U=U^Aq_A,
\end{equation}
as well as the (complex differential) eth operators $\eth$ and $\bar \eth$
~\cite{eth}.

Einstein's equations $R_{\alpha\beta}=8\pi(T_{\alpha\beta}
-\frac{1}{2}g_{\alpha\beta}T)$ are classified as: hypersurface
equations -- $R_{11},q^AR_{1A},h^{AB}R_{AB}$ -- forming a hierarchical
set for $\beta,U$ and $W_c$; evolution equation $q^Aq^B R_{AB}$ for
$J$; and constraints $R_{0\alpha}$. An evolution problem is normally
formulated in the region of spacetime between a timelike or null
worldtube and future null infinity, with (free) initial data
$J$ given on $u=0$, and with boundary data for $\beta,U,W_c,J$ satisfying
the constraints given on the inner worldtube.


\subsection{The spin-weighted formalism and the $\eth$ operator}
A complex dyad is written
\begin{equation}
q_A=(r_2 e^{i\phi_2},r_3 e^{i\phi_3})
\end{equation}
where $r_A, \phi_A$ are real quantities (but in general they are not
vectors). The real and imaginary parts of $q_A$ are
unit vectors that are orthogonal to each other, and $q_A$ represents the
metric. Thus
\begin{equation}
q^A q_A = 0, q^A \bar q_A = 2,
q_{AB}=\frac{1}{2}(q_A\bar{q}_B+\bar{q}_A q_B).
\label{eq:diad.norm.def}
\end{equation}
It is straightforward to substitute the a 2-metric into
Eq.~(\ref{eq:diad.norm.def}) to find $r_A$ and $(\phi_3-\phi_2)$.
Thus $q_A$ is not unique, up to a unitary factor: if $q_A$ represents
a given 2-metric, then so does $q^\prime_A=e^{i\alpha}q_A$. Thus,
considerations of simplicity are used in deciding the precise form of
dyad to represent a particular 2-metric. For example, the dyads
commonly used to represent some unit sphere metrics, namely
spherical polars and stereographic, are
\begin{equation}
ds^2=d\theta^2+\sin^2\theta^2d\phi^2: q_A=(1,i\sin\theta);\;\;
ds^2=\frac{4(dq^2+dp^2)}{(1+q^2+p^2)^2}: q_A=\frac{2}{1+q^2+p^2}(1,i).
\end{equation}

Having defined a dyad, we may construct complex quantities representing
all manner of tensorial objects, for example $X_1=T_A q^A$,
$X_2=T^{AB} q_A \bar{q}_B$, $X_3=T^{AB}_C \bar{q}_A\bar{q}_B\bar{q}^C$.
Each object has no free indices, and has associated with it a spin-weight
$s$ defined as the number of $q$ factors less the number of $\bar{q}$
factors in its definition. For example, $s(X_1)=1,s(X_2)=0,s(X_3)=-3$,
and, in general, $s(X)=-s(\bar{X})$.
We define derivative operators $\eth$ and $\bar{\eth}$ acting on
a quantity $V$ with spin-weight $s$
\begin{equation}
\eth V=q^A \partial_A V + s \Gamma V,\;\;
\bar{\eth} V=\bar{q}^A \partial_A V - s \bar{\Gamma} V
\end{equation}
where the spin-weights of $\eth V$ and $\bar{\eth} V$ are $s+1$ and $s-1$,
respectively, and where
\begin{equation}
\Gamma=-\frac{1}{2} q^A\bar{q}^B\nabla_A q_B.
\label{e-G}
\end{equation}
In the case of spherical polars, $\Gamma=-\cot\theta$, and for stereographic
coordinates $\Gamma=q+ip$.

The spin-weights of the quantities used in the Bondi-Sachs metric are
\begin{equation}
s(W_c)=s(\beta)=0,\; s(J)=2,\; s(\bar{J})=-2,\; s(U)=1,\; s(\bar{U})=-1.
\end{equation}

We will be using spin-weighted spherical harmonics~\cite{newp,
golm} using the formalism described in~\cite{mod}. It will
prove convenient to use ${}_s Z_{\ell m}$ rather than the usual
${}_s Y_{\ell m}$ (the suffix ${}_s$ denotes the spin-weight)
as basis functions, where
\begin{eqnarray}
{}_s Z_{\ell m} &=& \frac{1}{\sqrt{2}} \left[{}_s Y_{\ell m}
   +(-1)^m {}_s Y_{\ell -m}\right] \mbox{ for } m>0 \nonumber \\
{}_s Z_{\ell m} &=& \frac{i}{\sqrt{2}} \left[(-1)^m{}_s Y_{\ell m} 
   -{}_s Y_{\ell -m} \right]\mbox{ for }  m<0 \nonumber \\
{}_s Z_{\ell 0} &=& {}_s Y_{\ell 0},
\end{eqnarray}
and note that~\cite{mod} uses the notation ${}_s R_{\ell m}$
rather than the ${}_s Z_{\ell m}$ used here; we
use a different notation to avoid any confusion with the Ricci
tensor. In the case $s=0$, the $s$ will be omitted, i.e.
$Z_{\ell m}={}_0 Z_{\ell m}$.
Note that the effect of the $\eth$ operator acting on $Z_{\ell m}$ is
\begin{equation}
\eth Z_{\ell m}=\sqrt{\ell(\ell+1)}\;{}_1Z_{\ell m}, \;\;\;
\eth^2 Z_{\ell m}=\sqrt{(\ell -1)\ell(\ell+1)(\ell+2)}\;{}_2Z_{\ell m}.
\end{equation}


\subsection{Linearized solutions}
\label{s-LS}
A class of solutions, in Bondi-Sachs form, to the linearized Einstein
equations in vacuum was presented in ref.~\cite{BS-lin}, and we use these
solutions to test the accuracy of the numerical evolutions described
later. More specifically, the solutions to be used are those given in
Sec. 4.3 of ref.~\cite{BS-lin} for the case of a dynamic spacetime on
a Minkowski background. We write
\begin{eqnarray}
J&=& \sqrt{(\ell -1)\ell(\ell+1)(\ell+2)}\;{}_2Z_{\ell m} \Re(J_\ell(r) e^{i\nu u}), \;
U= \sqrt{\ell(\ell+1)}\;{}_1Z_{\ell m} \Re(U_\ell(r) e^{i\nu u}),
\nonumber \\
\beta&=& Z_{\ell m} \Re(\beta_\ell e^{i\nu u}),
\; W_c= Z_{\ell m} \Re(W_{c\ell}(r) e^{i\nu u}),
\label{e-an}
\end{eqnarray}
where $J_\ell(r)$, $U_\ell(r)$, $\beta_\ell$, $W_{c\ell}(r)$ are in general
complex, and taking the real part leads to $\cos(\nu u)$ and $\sin(\nu u)$
terms. The quantities $\beta$ and $W_c$ are real; while $J$ and $U$ are
complex due to the terms $\eth^2 Z_{\ell m}$ and  $\eth Z_{\ell m}$,
representing different terms in the angular part of the metric. We
require a solution that is well-behaved at future null infinity, and is
well-defined for $r \ge 2$, at which surface we set the inner boundary.
We find in the case $\ell=2$
\begin{eqnarray}
\beta_2&=&\beta_0 \nonumber \\
J_2(r)&=&\frac{24\beta_0 +3 i \nu C_1 - i \nu^3 C_2}{36}+\frac{C_1}{4 r}
       -\frac{C_2}{12 r^3} \nonumber \\
U_2(r)&=&\frac{-24i\nu \beta_0 +3 \nu^2 C_1 - \nu^4 C_2}{36} +\frac{2\beta_0}{r}
       +\frac{C_1}{2 r^2} +\frac{i\nu C_2}{3 r^3} +\frac{C_2}{4 r^4} \nonumber \\
W_{c2}(r)&=&\frac{24i\nu \beta_0 -3 \nu^2 C_1 + \nu^4 C_2}{6} 
           +\frac{3i\nu C_1 -6\beta_0-i\nu^3 C_2}{3r}
           -\frac{\nu^2 C_2}{r^2} +\frac{i\nu C_2}{r^3} +\frac{C_2}{2r^4},
\label{e-NBl2}
\end{eqnarray}
with the (complex) constants $\beta_0$, $C_1$ and $C_2$ freely specifiable.

We find in the case $\ell=3$
\begin{eqnarray}
\beta_2&=&\beta_0 \nonumber \\
J_3(r)&=&\frac{60\beta_0 +3 i \nu C_1 + \nu^4 C_2}{180}+\frac{C_1}{10 r}
       -\frac{i \nu C_2}{6 r^3} -\frac{C_2}{4r^4}
\nonumber \\
U_3(r)&=&\frac{-60i\nu \beta_0 +3 \nu^2 C_1 - i \nu^5 C_2}{180}
         +\frac{2\beta_0}{r} +\frac{C_1}{2 r^2}
         -\frac{2\nu^2 C_2}{3 r^3} +\frac{5 i \nu C_2}{4 r^4}
         + \frac{C_2}{r^5}\nonumber \\
W_{c3}(r)&=&\frac{60 i \nu \beta_0 -3 \nu^2 C_1 + i\nu^5 C_2}{15} 
           +\frac{i\nu C_1 -2\beta_0+\nu^4 C_2}{3r}
           -\frac{i2\nu^3 C_2}{r^2} -\frac{4i\nu^2 C_2}{r^3}
           +\frac{5\nu C_2}{r^4}+\frac{3 C_2}{r^5}.
\label{e-NBl3}
\end{eqnarray}

The emitted gravitational radiation, that is the news $N$, takes a
simple form in the linearized limit when the metric satisfies
Eq.~(\ref{e-an})
\begin{equation}
N=\Re\left(  e^{i\nu u}\lim_{r \rightarrow \infty}
\left(\frac{\ell(\ell+1)}{4}J_{\ell}-\frac{i\nu}{2} r^2 J_{\ell,r}\right)
 + e^{i\nu u}\beta_\ell \right) \sqrt{(\ell-1)\ell(\ell+1)(\ell+2)}
\;{}_2Z_{\ell m}.
\label{e-Nl}
\end{equation}
For the cases $\ell=$2 and 3,
\begin{equation}
\ell=2:\;\;N=\Re\left(\frac{i\nu^3 C_2}{\sqrt{24}} e^{i\nu u}\right)
  \;{}_2Z_{2m};\;\;\;
\ell=3:\;\;N=\Re\left(\frac{-\nu^4 C_2}{\sqrt{30}} e^{i\nu u}\right)
  \;{}_2Z_{3m}.
\label{e-N}
\end{equation}


\subsection{Six-patch Angular Coordinates}

Given Cartesian coordinates $(x,y,z)$, we define ``inflated-cube''
angular coordinates on each 2-sphere of constant $u$ and~$r$,
\begin{equation}
					           \label{eqn-6-patch-mu-nu-phi}
\renewcommand{\arraystretch}{1.25}
\begin{array}{c@{}l@{}l}
\mu	& {} \equiv \text{rotation angle about the $x$ axis}
	& {} = \arctan(y/z)						\\
\nu	& {} \equiv \text{rotation angle about the $y$ axis}
	& {} = \arctan(x/z)						\\
\phi	& {} \equiv \text{rotation angle about the $z$ axis}
	& {} = \arctan(y/x)						
\end{array}
\end{equation}
where all the arctangents are 4-quadrant based on the signs of $x$, $y$,
and $z$.  We then introduce 6~coordinate patches covering neighborhoods
of the $\pm z$, $\pm x$, and $\pm y$ axes, with the angular coordinates
$x^A \equiv (\rho,\sigma)$ in each patch defined as follows
(see note
\footnote{
         This definition differs from that of \cite{jthorn}
         only by the interchange of $\rho$ and $\sigma$ in
         $\pm z$ patches.  The present definition has the
         advantage of having a consistent parity, with
         $\vec{\rho} \times \vec{\sigma}$ always pointing
         $\{\text{outwards},\text{inwards}\}$ in $\{+,-\}$~patches.
         }
):
\begin{equation}
(\rho,\sigma) \equiv	\left\{
	\begin{array}{ll}
	(\nu,\mu)	& \text{in $\pm z$ patches}	\\
	(\nu,\phi)	& \text{in $\pm x$ patches}	\\
	(\mu,\phi)	& \text{in $\pm y$ patches}	
	\end{array}
	\right.
							 \label{eqn-6-patch-x^A}
\end{equation}

Notice that each patch's $x^A$ coordinates are nonsingular throughout
a neighborhood of the patch, and that the union of all the patches
covers $S^2$ without coordinate singularities.  The name ``inflated-cube''
comes from another way to visualize these patches and coordinates:
Imagine an $xyz$~cube with $xyz$~grid lines painted on its face.
Now imagine the cube to be flexible, and inflate it like a balloon,
so it becomes spherical in shape.  The resulting coordinate lines
will closely resemble those for $(\mu,\nu,\phi)$ coordinates.

\begin{figure}[tbp]
\begin{center}
\begin{picture}(125,125)
\put(26,-34){\includegraphics[scale=2.0,trim=45mm 20mm 20mm 40mm]{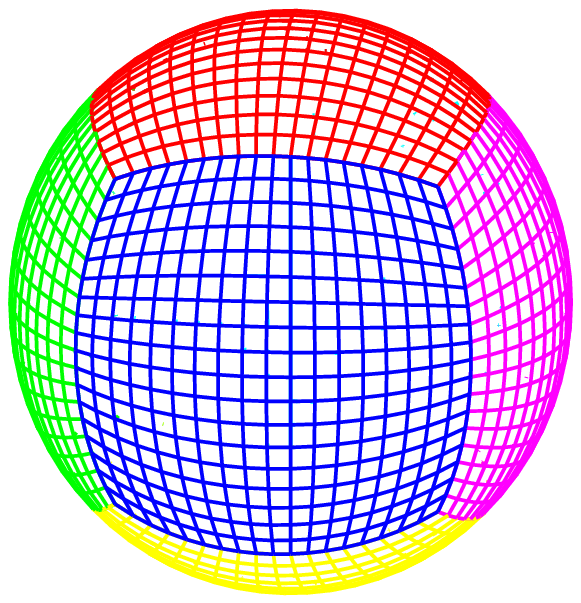}}
\put(26,-34){\includegraphics[scale=2.0,trim=45mm 20mm 20mm 40mm]{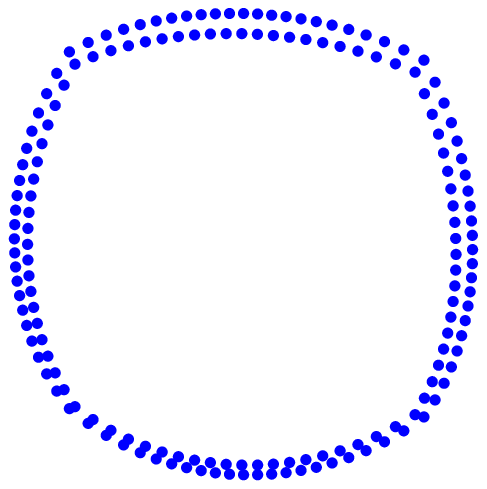}}
\end{picture}
\end{center}
\caption{
	This figure shows a six-patch grid covering $S^2$ at
	an angular resolution of $\Delta x^A = 5^\circ$.
	Each patch's nominal grid is shown with solid lines;
	the central patch's ghost-zone points are shown with
	solid dots.  The ghost zone has a width of 2~grid points
	(suitable for 5-point angular finite difference molecules).
	}
\label{fig-6-patch-with-ghost-zones}
\end{figure}

We introduce ghost zones in the usual manner along the angular boundaries
of each patch, and we refer to the non-ghost-zone part of each patch's
numerical grid as the ``nominal'' grid.  We size the patches so they
overlap slightly, with each ghost-zone grid point lying in the nominal
grid of some other patch.  Figure~\ref{fig-6-patch-with-ghost-zones}
shows an example of a six-patch system of this type.

We couple the patches together by interpolating the field variables
from neighboring patches to each ghost-zone grid point.  Notice that
with the definition~\eqref{eqn-6-patch-x^A}, the angular coordinate~$x^A$
perpendicular to an interpatch boundary is always common to both the
adjacent patches.  This means that the interpatch interpolation need
only be done in 1~dimension, parallel to the interpatch boundary.

Each patch uses its own $x^A\equiv(\rho,\sigma)$ dyad for the
spin-weighted quantities; when interpolating data from one patch to
another we transform the data (after interpolation) as described in
section~\ref{sect-spin-weighted-formalism}.


\section{Implementation of six patch angular coordinates}
\label{s-6patch}

\subsection{Spin-weighted formalism}
\label{sect-spin-weighted-formalism}

The unit sphere metric $q_{AB}$ in each patch is written as
\begin{equation}
ds^2=\left(1- \sin^2 \rho \; \sin^2 \sigma \right)^{-2} 
\bigg(
\cos^2\sigma  \; d\rho^2 + \cos^2\rho \;  d\sigma^2
- \frac{1}{2}  \sin(2\rho) \sin(2\sigma) \; d\rho \, d\sigma
\bigg),
\label{e-m1}
\end{equation}
with respect to coordinates $(\rho,\sigma)$ with range
$(-\pi/4,+\pi/4)\times (-\pi/4,+\pi/4)$.
A (simple) dyad representing Eq.~(\ref{e-m1}) is
\begin{equation}
q_A=\bigg(
\frac{(\theta_c+i \theta_s)\cos\sigma}{4\theta_c^2 \theta_s^2},
\frac{(\theta_c-i \theta_s)\cos\rho}{4\theta_c^2 \theta_s^2}
\bigg),\;\;
q^A=\bigg(
2\theta_c\theta_s\frac{\theta_s
+i \theta_c}{\cos\sigma},
2\theta_c\theta_s\frac{\theta_s
-i \theta_c}{\cos\rho}
\bigg),
\end{equation}
where
\begin{equation}
\theta_c=\sqrt{\frac{1-\sin\rho\sin\sigma}{2}},
\theta_s=\sqrt{\frac{1+\sin\rho\sin\sigma}{2}}.
\end{equation}
The factor $\Gamma$, defined in Eq.~(\ref{e-G}) and needed for the
evaluation of $\eth$, works out to be
\begin{eqnarray}
\Gamma&=&\frac{\cos^2\rho\cos^2\sigma(\sin\rho+\sin\sigma)
     +(\cos^2\rho-\cos^2\sigma)(\sin\sigma-\sin\rho)}
             {4 \theta_c \cos\sigma \cos\rho} \nonumber \\
     &+&i\frac{\cos^2\rho \cos^2\sigma (\sin\rho-\sin\sigma)
     +(\cos^2\sigma-\cos^2\rho)(\sin\rho+\sin\sigma)}
            {4 \theta_s \cos\sigma \cos\rho}.
\end{eqnarray}

In addition, we need to specify how spin-weighted quantities
transform at inter-patch boundaries.
Suppose that we have two patches ``Old'' and ``New'', with quantities
in the two patches being denoted by means of suffices ${}_{(O)}$ and
${}_{(N)}$, respectively. Define the Jacobian $J^A_B$ from the
old patch to the new patch
\begin{equation}
J^A_B=\frac{\partial x^A_{(N)}}{\partial x^B_{(O)}}.
\end{equation}
Then the dyad $q^A_{(O)}$ has components in the new coordinates
\begin{equation}
q^A_{(O)[N]}=q^B_{(O)} J^A_B
\end{equation}
where the notation is that the ${}_{(O)}$ indicates that we are
referring to the dyad that generates the metric in the old patch,
and the ${}_{[N]}$ indicates that the components are given with
respect to the new coordinates. Any two complex dyads are related
by means of a rotation $\exp(i\gamma)$. Writing
\begin{equation}
q^A_{(N)}=\exp(i\gamma) q^A_{(O)[N]},
\end{equation}
and applying Eq.~(\ref{eq:diad.norm.def}), it follows that
\begin{equation}
\exp(i\gamma)=\frac{2}{q_{AB(N)} \bar{q}^A_{(N)} q^B_{(O)[N]}},
\end{equation}
and then a spin-weighted quantity $V$ with spin-weight $s$ transforms
between the two patches as
\begin{equation}
V_{(N)}=\exp(i s \gamma) V_{(O)}.
\label{eqn-spin-transformation}
\end{equation}
There are many different cases for the transformation between
the patches, and we give
the transformation and Jacobian explicitly in only one case, when
the inter-patch boundary is at
\begin{equation}
\rho_{(O)}=\frac{\pi}{4},\; \rho_{(N)}=-\frac{\pi}{4},\;
\sigma_{(O)}=\sigma_{(N)}.
\end{equation}
Then the coordinate transformation is
\begin{equation}
\rho_{(N)}=\arctan\left(-\frac{1}{\tan \rho_{(O)}}\right),\;
\sigma_{(N)}=\arctan\left(\frac{\tan \sigma_{(O)}}
{\tan \rho_{(O)}}\right),
\end{equation}
and the Jacobian evaluates to
\begin{equation}
\frac{\partial \rho_{(N)}}{\partial \rho_{(O)}} =1,\;
\frac{\partial \rho_{(N)}}{\partial \sigma_{(O)}} =0,
\end{equation}
\begin{equation}
\frac{\partial \sigma_{(N)}}{\partial \rho_{(O)}} =
     -\frac{\cos\sigma \sin\sigma}
     {\cos^2\sigma \sin^2\rho+\cos^2\rho \sin^2\sigma},\;
\frac{\partial \sigma_{(N)}}{\partial \sigma_{(O)}}=
     \frac{\cos\rho \sin\rho}
     {\cos^2\sigma \sin^2\rho+\cos^2\rho \sin^2\sigma}.
\end{equation}


\subsection{Computational implementation}
The existing stereographic code has been extended to the six patch
coordinate system. Since the formulation of the equations in terms of
spin-weighted quantities is independent of angular coordinate bases but
dyad-dependent, it is necessary to re-implement only those objects that depend
on the six patch dyad. 
We have therefore provided a numerical implementation of the new
$\eth$-operators in the code. In addition, we have adapted the
spin-transformation coefficients to the six patches. 
In order to test the six-patch code against the linearized solutions,
the spin-weighted spherical harmonics ${}_sZ_{\ell m}$ needed to be 
implemented for the six-patch coordinates and dyad. 

The $\eth$-operators are implemented via subroutines $D_1(s, e)$ and
$D_2(s, e_1, e_2)$, which calculate $\eth$ or $\bar{\eth}$ and the
second derivative as the combinations $\eth\eth$, $\eth\bar{\eth}$,
$\bar{\eth}\eth$ and $\bar\eth\bar\eth$. The parameter $s$ specifies the spin
$s$ of the quantity to which the operator is applied, and $e_1,e_2\in\{-1,1\}$
denote $\eth$ and $\bar\eth$.

The derivative operators $\partial_\rho^2$, $\partial_\sigma^2$,
$\partial_\rho\partial_\sigma$, $\partial_\rho$ and $\partial_\sigma$ have been
approximated by second- and fourth-order accurate centered finite difference
stencils \cite{fornberg}. There is an input parameter that enables switching
between second- and fourth-order accurate derivatives.

In order to calculate the values of the $\eth$-derivatives at the boundary of
each patch, we need to access values from the neighboring patches. This is done
by defining ghost zones, which contain the needed values of these patches. The
multipatch infrastructure interpolates the spin-weighted (scalar) quantities
between the patches using either cubic or quintic Lagrange polynomial
interpolation, depending on whether angular derivatives are approximated
by second- or fourth-order accurate finite differences. In both cases,
the ghost-zone points lie on grid lines, so the interpolation was simple
to implement because it was only one dimensional.  After the interpolation,
we apply the transformation law~\eqref{eqn-spin-transformation} to transform
quantities of spin-weight $s\neq0$ to the current patch's coordinates and dyad.

It turns out that we need a total number of $12$ different spin-transformation
coefficients, since we have a $P_\pm$-symmetry between the total number of $24$
ghost zones across all $P_{+i}$ and $P_{-i}$, $i=x,y,z$ patches.
These coefficients are calculated and stored for repeated use in an initial
routine. After each radial step during the integration of the characteristic
equations, the ghost zones are synchronized by the multipatch infrastructure,
and afterwards, the code multiplies the appropriate spin-transformation
coefficients with the synchronized values of the ghost zones.

We have carried out checks of the angular grid including the $D_1$- and
$D_2$-operators and the spin-transformation coefficients, and we have found out
that the code converges with second- and fourth-order accuracy, respectively.

The spin-weighted spherical harmonics have been implemented by using the 
spherical harmonics in terms of the stereographic coordinate $\zeta=p+iq$
and by
applying the pseudo-numerical operators $D_1$ and $D_2$ in order to
obtain ${}_sZ_{\ell m}$ for
$|s|>0$. With pseudo-numerical, we mean that we apply the
fourth-order $D_1$ and $D_2$ operators with a very
small delta-spacing such that we reach machine precision, since we know the
$Z_{\ell m}$ everywhere and 
are not bound to the numerical grid. 
In order to use the stereographic routines for the $Z_{\ell m}$,
we have transformed
the stereographic coordinate $\zeta$ to the six-patch coordinates
and depending on the hemisphere, we use the stereographic routines 
for north or south patch, respectively.

Furthermore, we have implemented an algorithm for
calculating the linearized news function (Eq.(\ref{e-Nl})), in the form
\begin{equation}
N=\lim_{r \rightarrow \infty}
\left(\frac{\ell(\ell+1)}{4}J-\frac{1}{2} r^2 J_{,ur}
 + \eth^2\beta \right).
\end{equation}


\section{The constraint equations}
\label{s-cons}
If the boundary data satisfies the constraints $R_{0\alpha}=0$ (here we
restrict attention to the vacuum case), and provided the hypersurface
and evolution equations are satisfied,
the Bianchi identities guarantee that the constraints are satisfied
throughout the domain~\cite{bondi,sachs}. Thus from an analytic viewpoint,
evaluation of the constraints is redundant, but in a numerical simulation
their evaluation may provide useful information concerning the reliability
of the computation.

We have written code that uses the Bondi-Sachs metric variables and
derivatives to evaluate the following quantities
\begin{equation}
R_{00}, \;\; R_{01} \mbox{ and } q^A R_{0A}.
\end{equation}
The expressions for the above quantities are very long and are not reproduced
here. The Fortran code for these expressions was generated directly from
the computer algebra (Maple) output.


\section{Testing the code}
\label{s-res}
In this section we first specify the linearized solutions against which the
code will be tested, as well as the various parameters that describe a
numerical solution and its output. Then we present, as Figures and Tables,
the results of testing the
comparative performance of the second order six-patch, fourth order
six-patch and stereographic codes.

We refer to the linearized solutions summarized in Sec.~\ref{s-LS}. In all
cases we take
\begin{equation}
\nu=1 \mbox{ and } m=0.
\end{equation}
We present results for the cases $\ell=2$ and $\ell=3$ with
\begin{eqnarray}
C_1=3\cdot10^{-6}, \qquad C_2=10^{-6}, \qquad \beta_0=i \cdot 10^{-6} \qquad (\ell=2) \\
C_1=3\cdot10^{-6}, \qquad C_2=i \cdot 10^{-6}, \qquad \beta_0=i \cdot 10^{-6} \qquad (\ell=3)
\end{eqnarray}
in Eq. (\ref{e-NBl2}) in the case $\ell=2$, and in Eq. (\ref{e-NBl3}) in the
case $\ell=3$.

All the numerical simulations use a compactified radial coordinate $x=r/(r_{wt}+r)$
with $r_{wt}=9$. Data is prescribed at time $u=0$ as well as at the inner boundary $r=2$
(which is equivalent to $x=0.1888$). The stereographic grids (with ghost zones excluded)
are
\begin{equation}
\mbox{Coarse: } n_x=n_q=n_p=41, \qquad \mbox{ Fine: } n_x=n_q=n_p=81;
\end{equation}
and there is no overlap between the two patches, i.e. we set the code parameter $q_{\mbox{size}}=1$ which 
means that on the nominal grid, the holomorphic coordinate function $\zeta=q+ip$ takes values in $q,p\in [-1,1]$.
The six-patch grids are such that, over the whole sphere, the total number of angular
cells is equivalent. We take
\begin{equation}
\mbox{Coarse: } n_x=41, \quad n_\sigma=n_\rho=24, \qquad \mbox{ Fine: }
n_x=81, \quad n_\sigma=n_\rho=47.
\end{equation}
Six-patch results are reported for both second-order and fourth order differencing
of the angular derivatives.
In all cases, the fine grid has $\Delta u = 0.0125$ and the coarse grid has
$\Delta u =0.025$. Runs are performed for two complete periods, i.e. starting at
$u=0$ and ending at $u=4\pi$.

Results are reported for the errors of the quantities shown using the $L_2$ norm,
evaluated at the time shown, averaged over all non-ghost grid-points over the whole sphere
and between the inner boundary and future null infinity. The norm of the
error in the news is averaged over the whole sphere at future null infinity.

The error $\epsilon$ for a quantity $\Psi$ is defined as
\begin{equation}
\epsilon (t) =\Vert\Psi_{numeric}-\Psi_{analytic}\Vert.
\end{equation}
The convergence factor $C$ is then defined as the ratio between the error $\epsilon$ of low and high resolution
\begin{equation}
C (t) =\frac{\epsilon(t)_{low}}{\epsilon(t)_{high}}.
\end{equation}

\begin{figure}[h!]
\centering
\resizebox{14cm}{!}
{
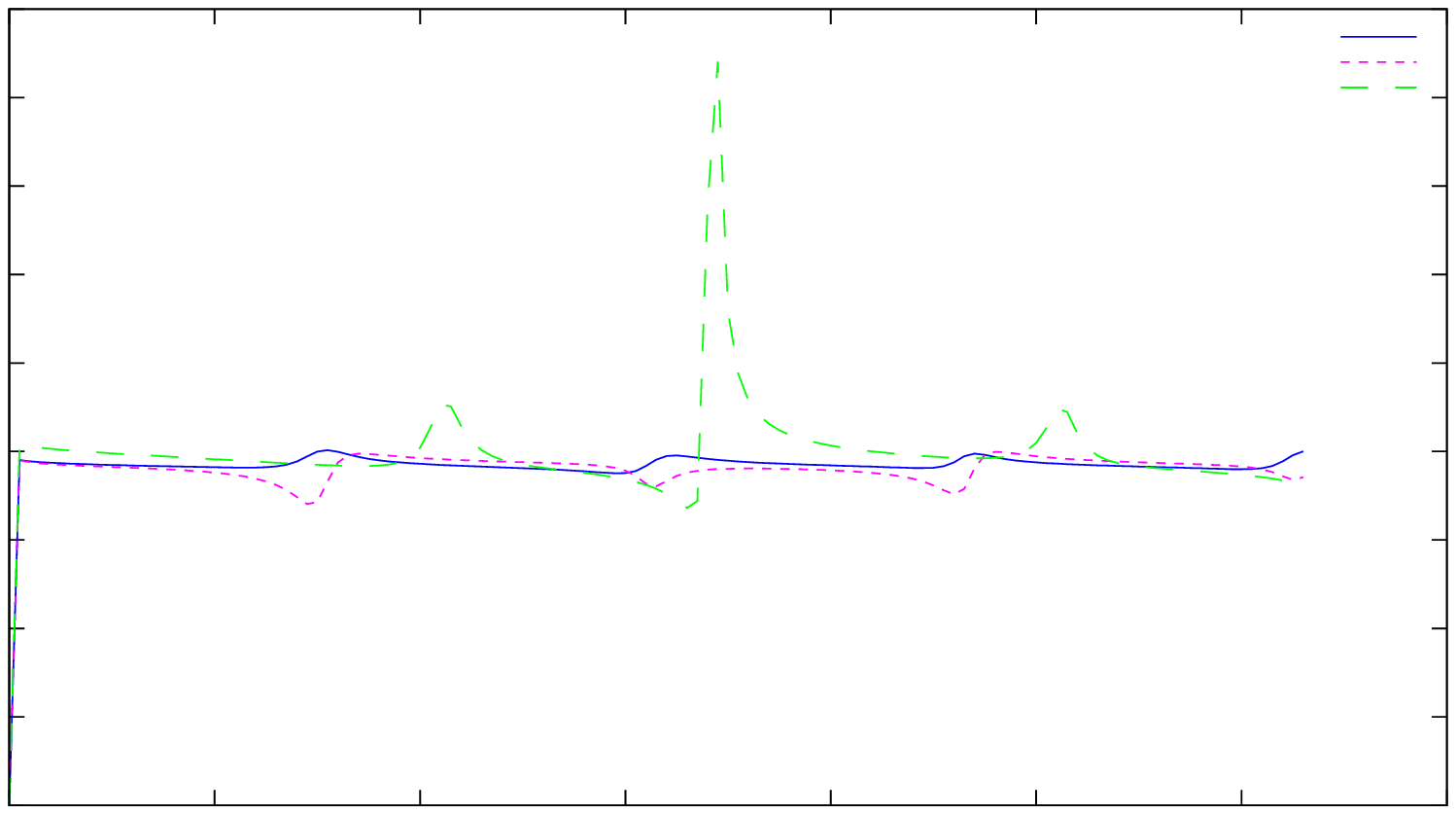
}\hfill

\resizebox{14cm}{!}
{
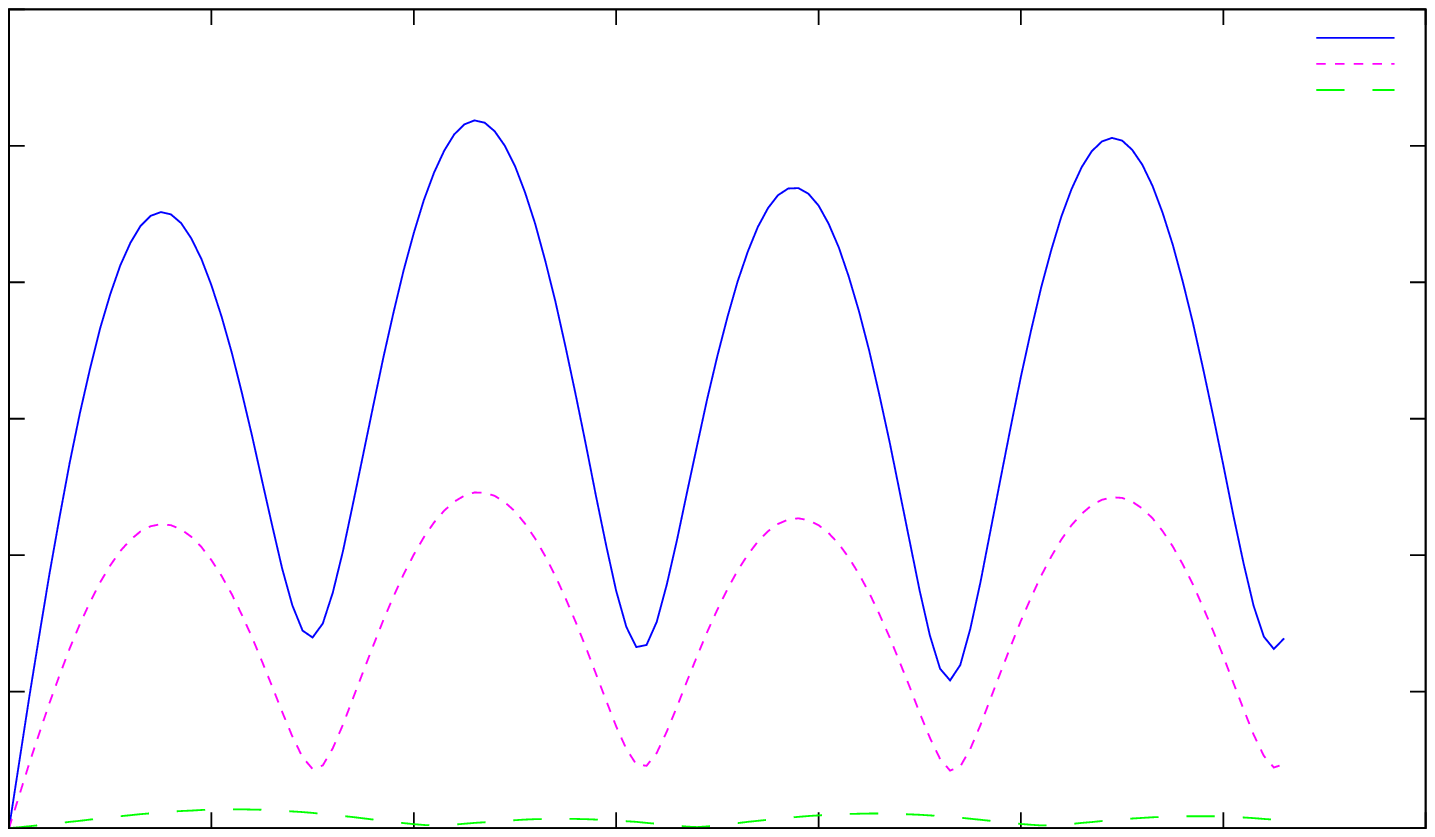
}\hfill
\caption{Convergence factor $C$ and error norm $\epsilon$ of $J$ plotted
against time in the case $\ell=2$}
\label{fig:nonBondi_l2J}
\end{figure}

\begin{figure}[h!]
\centering
\resizebox{14cm}{!}
{
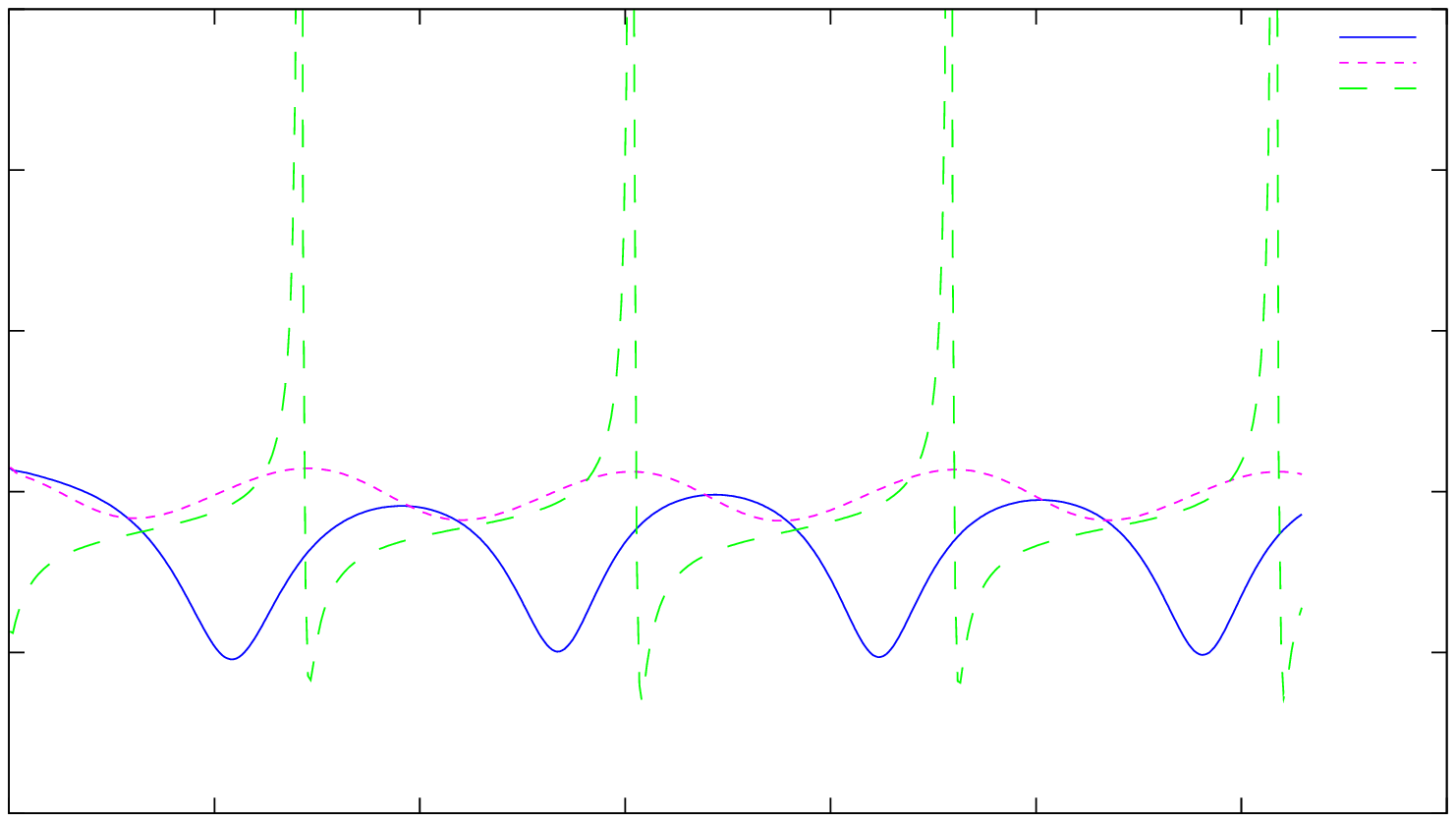
}\hfill

\resizebox{14cm}{!}
{
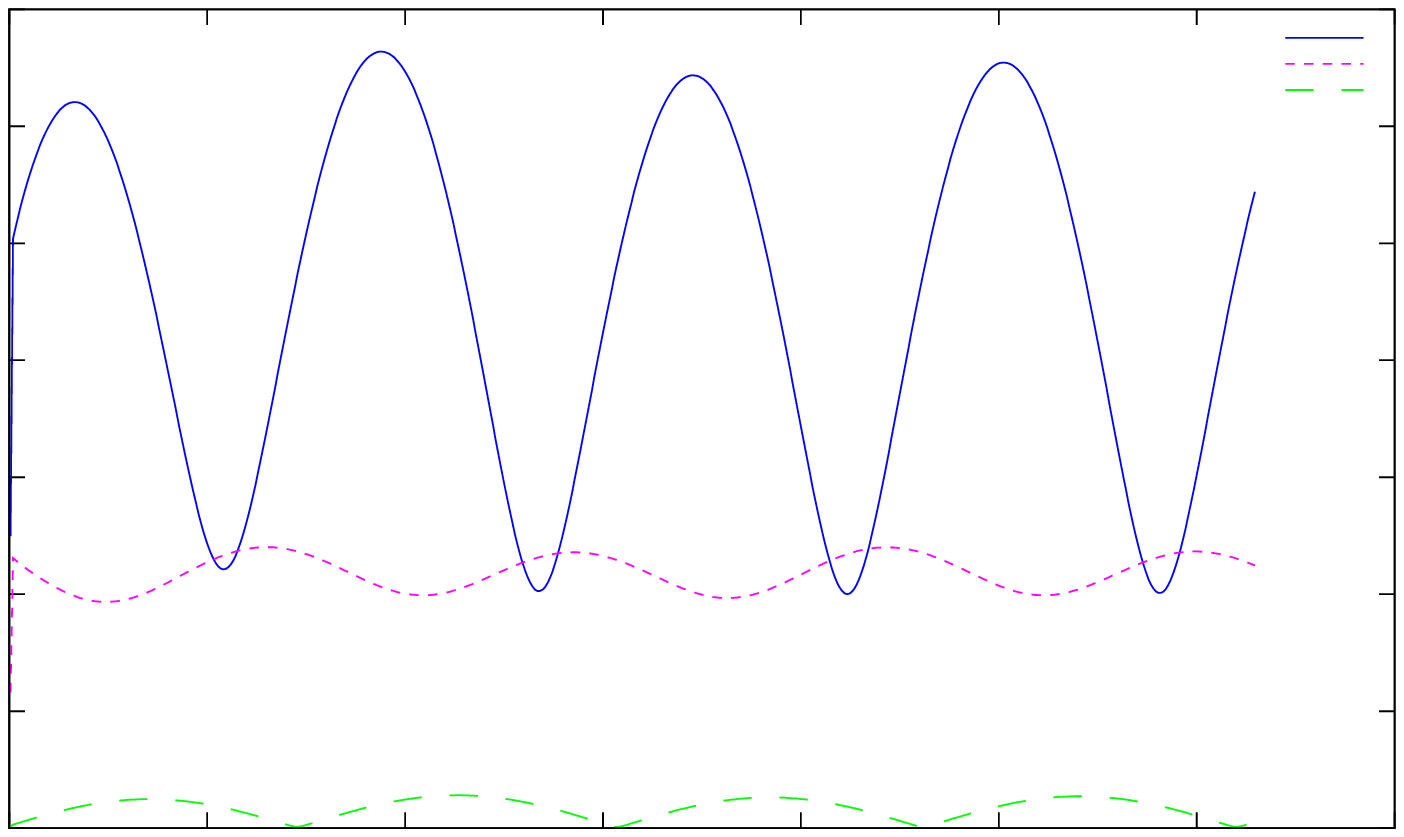
}\hfill
\caption{Convergence factor $C$ and error norm $\epsilon$ of $N$ plotted
against time in the case $\ell=2$}
\label{fig:nonBondi_l2N}
\end{figure}

\begin{figure}[h!]
\centering
\resizebox{14cm}{!}
{
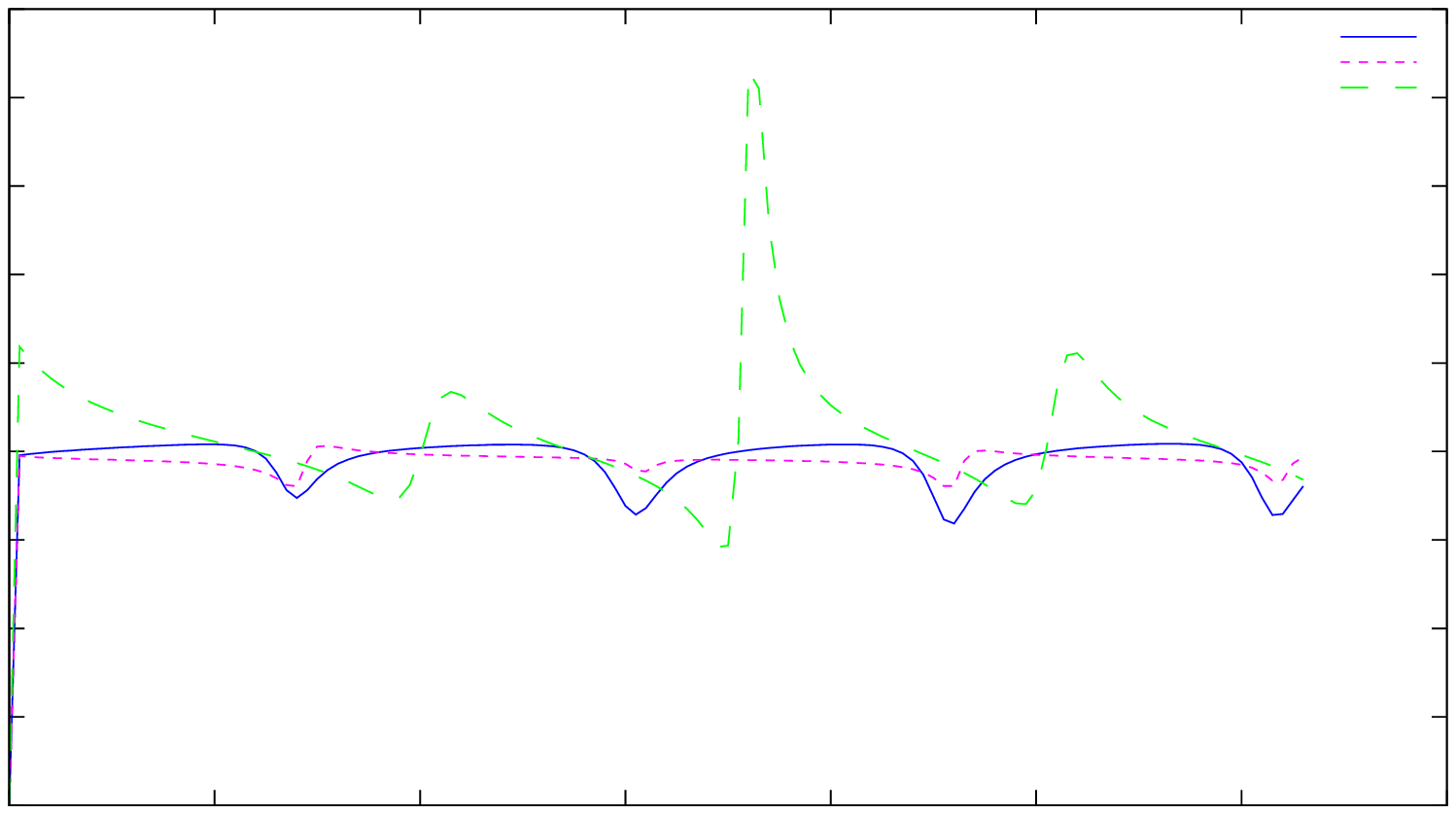
}\hfill

\resizebox{14cm}{!}
{
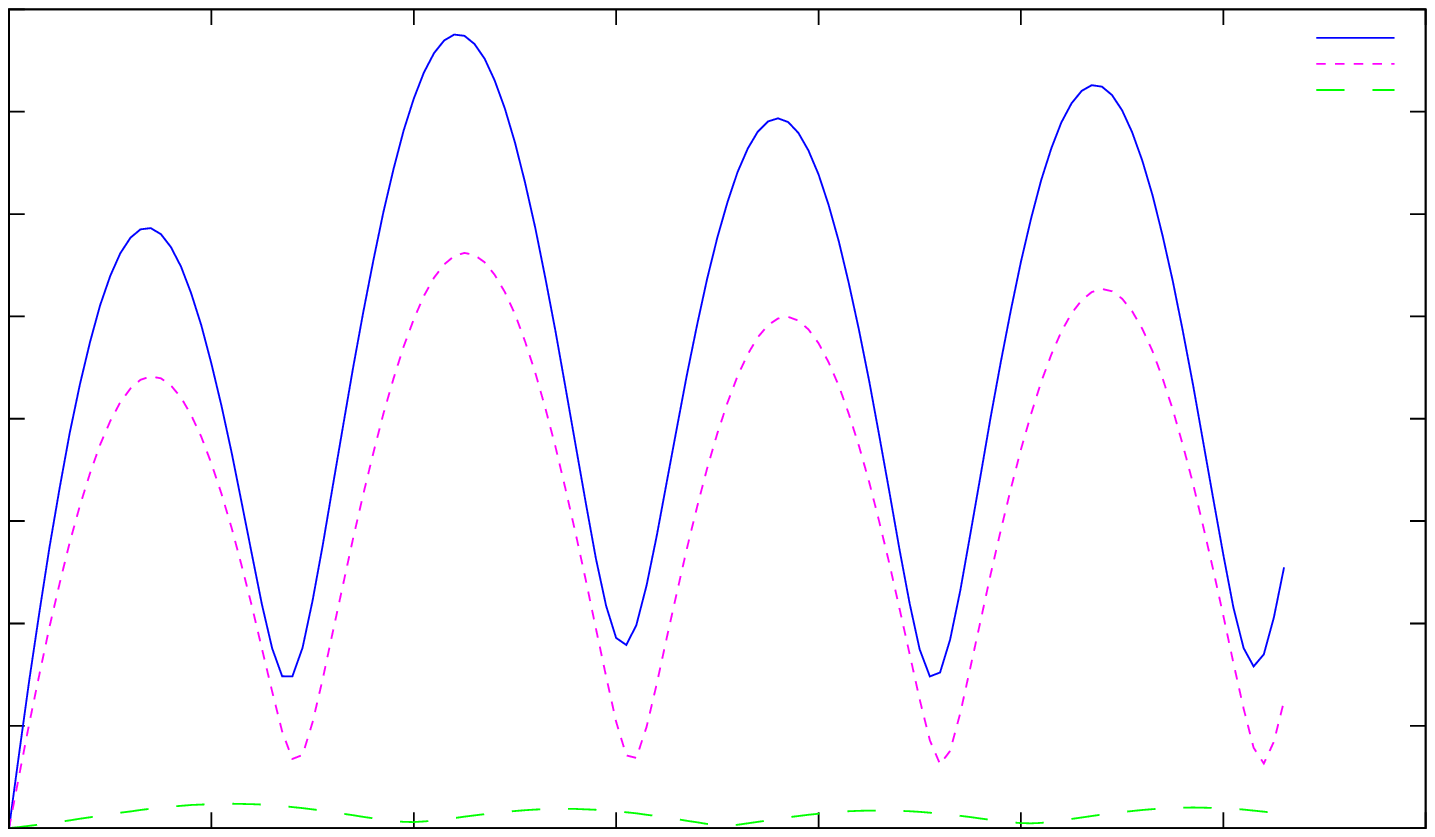
}\hfill
\caption{Convergence factor $C$ and error norm $\epsilon$ of $J$ plotted
against time in the case $\ell=3$}
\label{fig:nonBondi_l3J}
\end{figure}

\begin{figure}[h!]
\centering
\resizebox{14cm}{!}
{
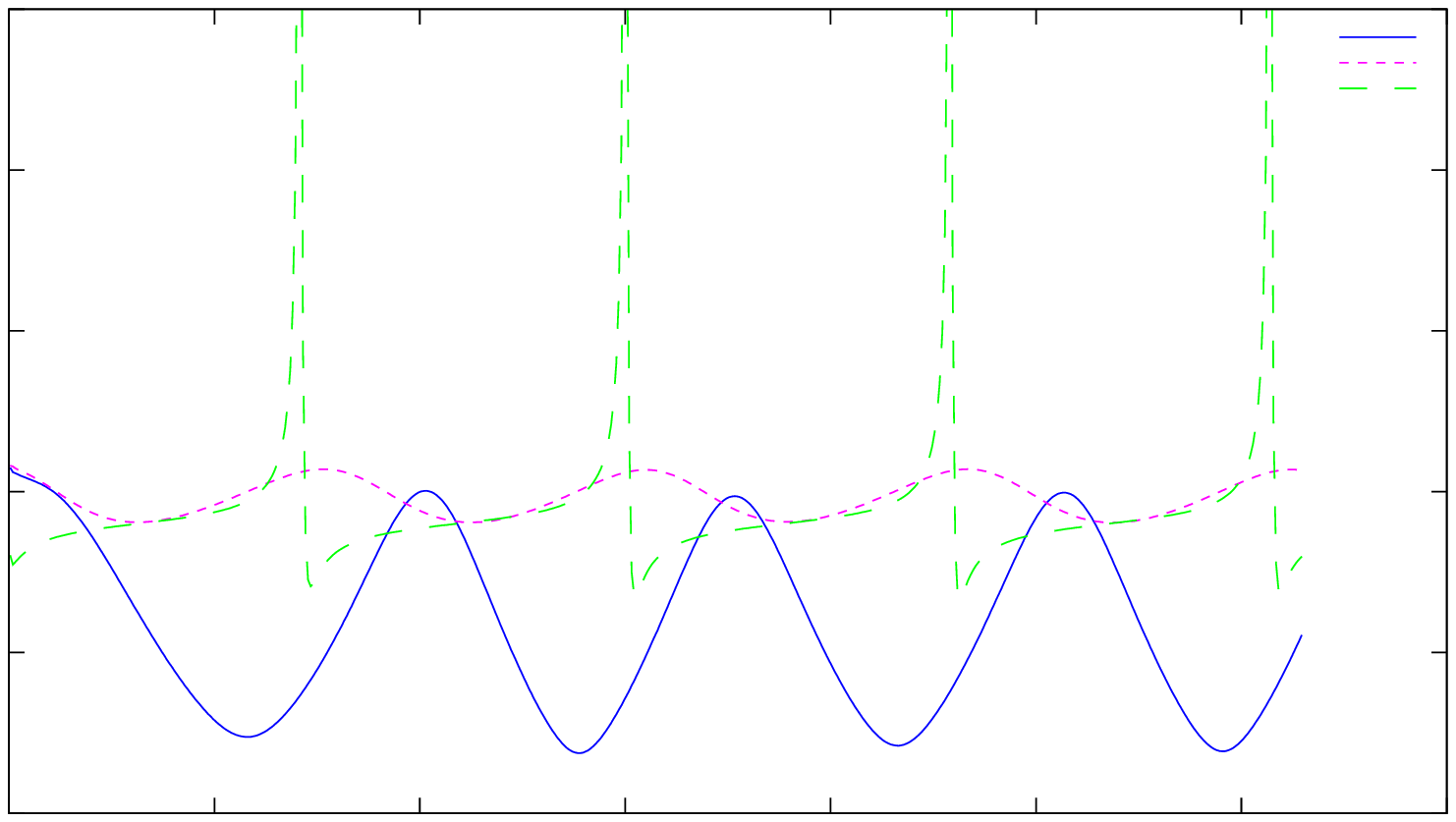
}\hfill

\resizebox{14cm}{!}
{
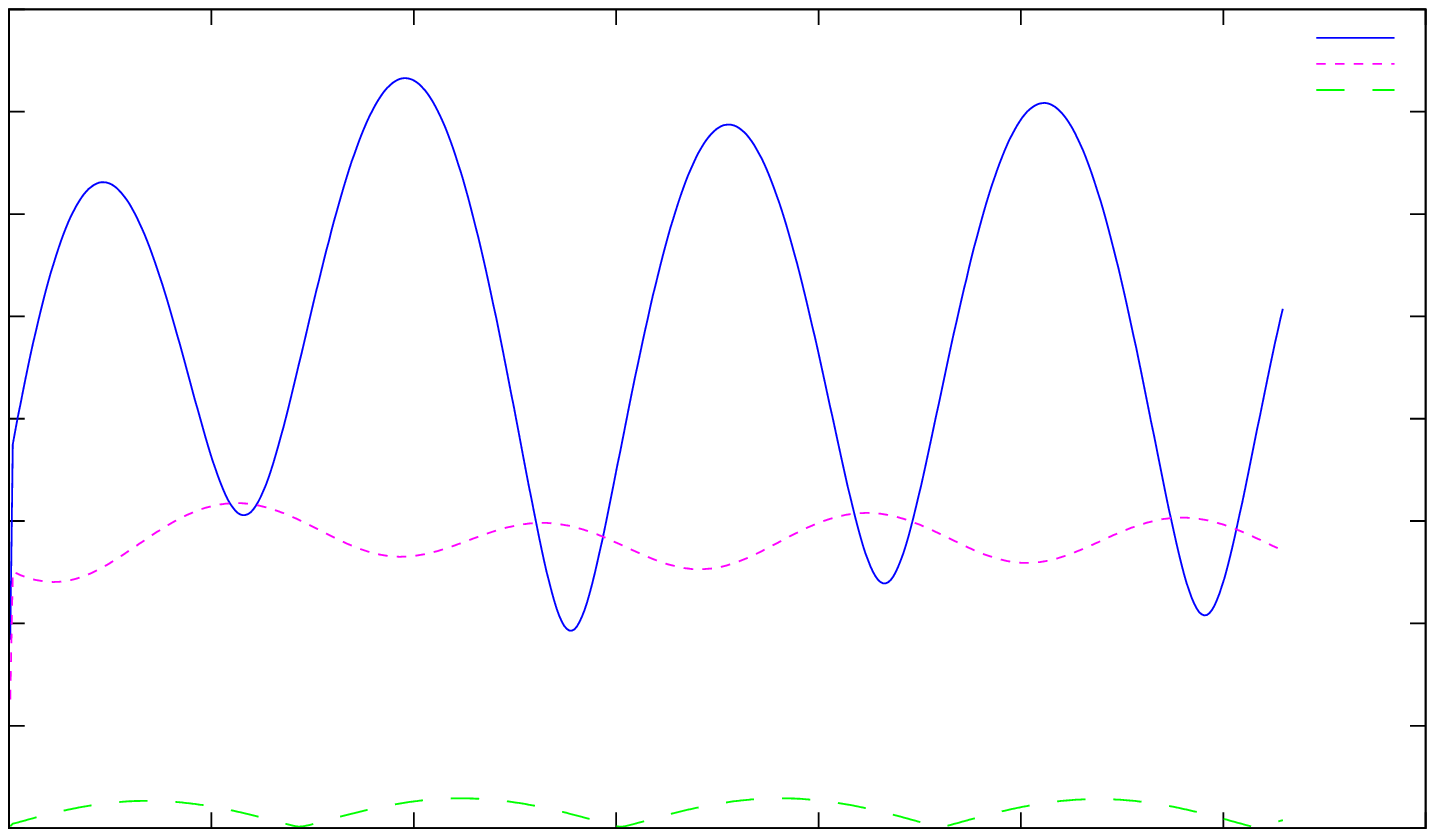
}\hfill
\caption{Convergence factor $C$ and error norm $\epsilon$ of $N$
plotted against time in the case $\ell=3$}
\label{fig:nonBondi_l3N}
\end{figure}

\begin{table}
\caption{Comparative performance in the $\ell=2$ case}
\begin{ruledtabular}
\begin{tabular}{lccc}
  & Stereographic & six-patch, 2nd order  & six-patch, 4th order \\
Averaged convergence rate of $J$ 
  & 3.8456 &  3.8286 &  3.9112 \\
Averaged absolute error of $J$ 
  & 3.3039$\times 10^{-9}$ & 1.5491$\times 10^{-9}$ & 6.9157$\times 10^{-11}$ \\
Averaged convergence rate of $N$ 
  & 3.3119 & 3.9642  &  3.5528 \\
Averaged absolute error of $N$
  & 2.2785$\times 10^{-8}$ & 1.0913$\times 10^{-8}$ & 8.4414$\times 10^{-10}$\\
Averaged convergence rate of $R_{00}$ 
  &  1.2487 & 1.5000 &  2.0319 \\
Averaged absolute error of $R_{00}$
  & 3.1942$\times 10^{-9}$ & 2.8779$\times 10^{-9}$ & 5.7668$\times 10^{-10}$\\
Averaged convergence rate of $R_{01}$ 
  & 3.5560  &  3.5936 &  3.1296\\
Averaged absolute error of $R_{01}$
  & 3.9214$\times 10^{-11}$ & 1.6988$\times 10^{-11}$ & 2.7331$\times 10^{-12}$ \\
Averaged convergence rate of $R_{0A}$ 
  & 3.4285 &  1.7558 &  2.0043 \\
Averaged absolute error of $R_{0A}$
  & 5.2549$\times 10^{-9}$ & 6.6397$\times 10^{-9}$ & 2.1543$\times 10^{-9}$
\end{tabular}  \end{ruledtabular}
\label{tab:NB2}
\end{table}

\begin{table}
\caption{Comparative performance in the $\ell=3$ case}
\begin{ruledtabular}
\begin{tabular}{lccc}
  & Stereographic & six-patch, 2nd order  & six-patch, 4th order \\
Averaged convergence rate of $J$ 
  & 3.9783 & 3.9106 &  4.0777 \\
Averaged absolute error of $J$ 
  & 4.6461$\times 10^{-9}$ & 3.2784$\times 10^{-9}$ &  1.3677$\times 10^{-10}$ \\
Averaged convergence rate of $N$ 
  &  2.1201 & 3.9134  &  3.6262\\
Averaged absolute error of $N$
  & 4.9174$\times 10^{-8}$ & 2.8182$\times 10^{-8}$ & 1.7996$\times 10^{-9}$\\
Averaged convergence rate of $R_{00}$ 
  & 1.2743  &  1.7963 &  2.0330 \\
Averaged absolute error of $R_{00}$
  & 7.2594$\times 10^{-9}$ & 4.5824$\times 10^{-9}$ & 1.1744$\times 10^{-9}$\\
Averaged convergence rate of $R_{01}$ 
  &  3.5144 & 3.5383  & 3.3824\\
Averaged absolute error of $R_{01}$
  & 1.3262$\times 10^{-10}$ & 7.5501$\times 10^{-11}$ & 6.0924$\times 10^{-12}$\\
Averaged convergence rate of $R_{0A}$ 
  & 3.4326  & 1.9510  & 2.0156 \\
Averaged absolute error of $R_{0A}$
  & 9.0299$\times 10^{-9}$ & 1.0076$\times 10^{-8}$ & 2.9654$\times 10^{-9}$
\end{tabular}  \end{ruledtabular}
\label{tab:NB3}
\end{table}


\section{Conclusion}
\label{s-conc}
We have implemented a version of the characteristic numerical relativity
code that coordinatizes the sphere by means of six angular patches. Further,
the six-patch code has been implemented for both second-order and
fourth-order accurate finite differencing of angular derivatives.

We compared the errors in the second-order six-patch, fourth-order
six-patch and stereographic versions of the code, using exact solutions of
the linearized Einstein equations as a testbed. This was done for a variety
of cases and using several different indicators to measure the error.
The convergence rate of the metric (i.e. of $J$) was always approximately
second-order, but in some cases we observed degradation of the order of
convergence of other quantities, all of which contain second derivatives of
the metric. This has shown up in previous performed runs of the code and
might be related to high-frequency error modes coming from angular patch
interfaces or corners. On average, the error norm of second order six-patch
was smaller than that of stereographic
by a factor of order two (although there were cases in which the error
was slightly larger). However, the fourth-order six-patch scheme
exhibited a dramatic reduction in the error norm, by a factor of up to 47
compared to that of the stereographic case.

Thus, we expect the six-patch characteristic code, in particular the
version that uses
fourth-order accurate angular finite differencing, to give significantly
better performance than the stereographic version.


\section*{Acknowledgments}
NTB and CWL thank Max-Planck-Institut f\"ur
Gravitationsphysik, Albert-Einstein-Institut, for hospitality;
and BS and CR thank the University of South Africa for hospitality. 
The work was supported in part by the National Research Foundation,
South Africa, under Grant number 2053724.

\end{document}